\documentclass[preprint,prc,aps,showpacs,superscriptaddress]{revtex4}
\usepackage{graphicx}
\usepackage{xcolor}

\usepackage{amsmath} 
\usepackage{amssymb} 

\newcommand{\bra}[1]{\langle #1|}
\newcommand{\ket}[1]{|#1\rangle}
\newcommand{\braket}[2]{\langle #1|#2\rangle}

\begin{document}
\title{The $g_{J/\psi D_s D_s}$ strong coupling constant from QCD Sum Rules}

\author{B. Os\'orio Rodrigues}
\affiliation{Instituto de F\'{\i}sica, Universidade do Estado do Rio de 
Janeiro, Rua S\~ao Francisco Xavier 524, 20550-900, Rio de Janeiro, RJ, Brazil. }

\author{M. E. Bracco}\email{bracco@uerj.br}
\affiliation{Faculdade de Tecnologia, Universidade do Estado do Rio de Janeiro, 
Rod. Presidente Dutra Km 298, P\'olo Industrial, 27537-000, Resende, RJ, Brazil.}

\author{M. Chiapparini}
\affiliation{Instituto de F\'{\i}sica, Universidade do Estado do Rio de 
Janeiro, Rua S\~ao Francisco Xavier 524, 20550-900, Rio de Janeiro, RJ, Brazil. }

\begin{abstract}
The coupling constant of the meson vertex $J/\psi D_s D_s$ is calculated using
the three point correlation function within the QCD Sum Rule formalism.
We have considered alternately mesons $J/\psi$ and $D_s$ off-shell together with non-perturbative contributions up to the mixed quark-gluon condensates. When extrapolated, these two different form factors give the same coupling constant 
for the process  $g_{J/\psi D_s D_s} = 5.98^{+0.67}_{-0.58}$.
\end{abstract}


\maketitle

\section{Introduction}
	In recent years, many charmonium-like states have been discovered, some
	of which do not easily fit into the conventional quark model picture.  
	 The Belle Collaboration observed an unexpected structure in the $\phi J/\psi$ mass
 which is the charmonium state  $X(4350)$ \cite{Shen:2010} and no significant
  signal was founded for the $Y(4140)$ as well. It does not contradict the CDF Collaboration
  measurement, which reported evidence of this state in the exclusive $B^+$ decay \cite{Aaltonen:2009}.
   These new hadrons, coming from different experiments \cite{Shen:2010,Aaltonen:2009},
   have produced many alternative theoretical interpretations (\cite{Mahajan:2009pj,Branz:2009yt,Liu:2009iw,Liu:2009ei,Zhao:2011sd}), which were proposed in order to better understand their  properties \cite{Shen:2012}. In some of them, the coupling constant of the $J/\psi D_s D_s$ vertex is a necessary input to calculate the strong decay. For example, in the works of Ref. \cite{Liu:2009iw}, decay $Y(4140) \to J/\psi \phi$ is studied  with the intermediate decays $Y(4140) \to D_s \bar{D}^{(*)}_s \to J/\psi \phi$, and Ref. \cite{Zhao:2011sd}, decay $X(4350) \to J/\psi \phi$ is analyzed using the intermediate process $X(4350) \to D_s^{(*)} \bar{D}_s^{(*)}  \to J/\psi \phi$.
 
 The strong coupling constant among these mesons can not be explained by perturbative
 theories, because the associate decays lie in the low energy region.  In the hadronic scale
 the QCD Sum Rule (QCDSR) approach is one of the most powerful non-perturbative methods, 
 which is also independent of model parameters. Not only to follow, but also to describe the procedure 
 in the experiment, as well as match it with the theoretical interpretation proposed, it is fundamental to know about the coupling constants and form factors
 among the decay process.
 
In this work we use the QCDSR formalism (following the development of previous works of our group (\cite{Bracco:2004rx,Rodrigues:2010ed,Bracco:2011pg}) to obtain the coupling constant and form factors of the $J/\psi D_s D_s$ vertex. 

Also, it is worth mentioning that a more precise knowledge  of this coupling constant and form factors will certainly improve the understanding of the fundamental constitution of these new observed mesons.

\section{Formalism}
The starting point for a three meson vertex calculation in the QCDSR formalism is the three point correlation function \cite{Bracco:2011pg}. Performing the sum rule, we are able to obtain  a form factor as a function of the momentum of one of the three mesons of the vertex. 

The coupling constant is obtained by extrapolating the form factor in the space-like region 
to $Q^2=-m^2$, where $m$ is the off-shell meson mass. In order to minimize the uncertainties 
associated with the extrapolation procedure, we perform the calculation twice,  
putting each meson off-shell -- usually the ones with the biggest mass difference -- obtaining two different form factors. Then it is required that these functions have the same value at their poles, extracting an unique coupling constant for the vertex. This procedure was introduced by our group many years ago (see for example \cite{Bracco:2011pg} and references therein). Note that the form factors are a side product of the coupling constant calculation within the QCDSR formalism, and they are of considerable interest in experimental high-energy physics.

In this work, we are interested in the $J/\psi D_s D_s$ vertex. Therefore, and as explained before, we have two different correlation functions to work with, one with the vector meson off-shell ($J/\psi$) and another with the pseudo-scalar meson ($D_s$) off-shell:
\begin{eqnarray}
\Gamma^{(J/\psi)}_\mu(p,p') &=& \int \bra{0'} T\{ j^{D_s}_5(x)j^{J/\psi\dagger}_\mu(y)j^{D_s\dagger}_5(0) \}\ket{0'} e^{ip'x}e^{-iqy}d^4x d^4y\,,  \label{eq:pivoff} \\
\Gamma^{(D_s)}_\mu(p,p') &= &\int \bra{0'} T\{ j^{D_s}_5(x)j^{D_s\dagger}_5(y)j^{J/\psi\dagger}_\mu(0) \}\ket{0'} e^{ip'x}e^{-iqy}d^4x d^4y\,, \label{eq:pidsoff}
\end{eqnarray}
where $q = p' - p$ is the transferred momentum. Eq.~(\ref{eq:pivoff}) is for
meson ${J/\psi}$ off-shell, which gives the form factor $g^{(J/\psi)}_{J/\psi D_s D_s}(q^2)$, and Eq.~(\ref{eq:pidsoff}) is for meson ${D_s}$ off-shell, giving the form factor $g^{(D_s)}_{J/\psi D_s D_s}(q^2)$, as we will see below.

According to the QCDSR approach, we can calculate these correlation functions in two different ways: either with hadron degrees of freedom (called the \textit{phenomenological side}) or with quark degrees of freedom (called the \textit{OPE side}). Both representations are equivalent -- invoking the quark-hadron duality -- and they are equated after applying a double Borel transform, when the form factors and coupling constant can be extracted.

\subsection{The phenomenological side}
In order to perform the calculation of the phenomenological side, it is necessary to know the effective Lagrangian of the interaction for the process ($J/\psi D_s D_s$). We choose a SU(4) 
Lagrangian, which is \cite{Liu:2009iw,Zhao:2011sd}: 
\begin{align}
\mathcal{L}_{J/\psi D_s D_s} = i g_{J/\psi D_s D_s} \psi^\alpha ( D^+_s\partial_\alpha D^-_s - \partial_\alpha D^+_s D_s^-)\,.
\label{eq:lagrangeana}
\end{align}

From this Lagrangian, we can obtain the vertex of the hadronic process. In the case of  $J/\psi$ off-shell we have:
\begin{eqnarray}
\braket{D_s(p)J/\psi(q)}{D_s(p')} &=& i g^{(J/\psi)}_{J/\psi D_s D_s}(q^2)\epsilon^\alpha(q)(p_\alpha + p'_\alpha) \,.
\end{eqnarray}
We also make use of the following hadronic matrix elements:
\begin{eqnarray}
\bra{0} j_5^{D_s} \ket{D_s(p')} &=& \bra{D_s(p)} j_5^{D_s} \ket{0} = f_{D_s} \frac{m^2_{D_s}}{m_c + m_s}\,,\\
\bra{J/\psi(q)} j_\mu^{J/\psi} \ket{0} &=& f_{J/\psi} m_{J/\psi} \epsilon^*_\mu(q)\,.
\end{eqnarray}
In the case of  $D_s$ off-shell the obtained vertex is: 
\begin{eqnarray}
\braket{D_s(q)J/\psi(p)}{D_s(p')} &=& i g^{(D_s)}_{J/\psi D_s D_s}(q^2)\epsilon^\alpha(p)(2p'_\alpha - p_\alpha)\,,
\end{eqnarray}
with the corresponding matrix elements:
\begin{eqnarray}
\bra{0} j_5^{D_s} \ket{D_s(p')} &=& \bra{D_s(q)} j_5^{D_s} \ket{0} = f_{D_s} \frac{m^2_{D_s}}{m_c + m_s}\,,\\
\bra{J/\psi(p)} j_\mu^{J/\psi} \ket{0} &=& f_{J/\psi} m_{J/\psi} \epsilon^*_\mu(p)\,.
\end{eqnarray}
In these formulas, $f_{J/\psi}$ and $f_{D_s}$ are the decay constants of $J/\psi$ and $D_s$ mesons respectively,  $\epsilon_{\mu}$ is the polarization vector, $g^{(J/\psi)}_{J/\psi D_s D_s}(q^2)$ and $g^{(D_s)}_{J/\psi D_s D_s}(q^2)$ are the form factors of the vertex (with mesons $J/\psi$ or $D_s$ off-shell respectively), and $m_s$ and $m_c$ are the strange and 
charm quark masses.

After the calculation we obtain the correlation functions for the phenomenological side in the form:
\begin{eqnarray}
\Gamma^{phen (J/\psi)}_\mu(p,p') &=& -\frac{g^{(J/\psi)}_{J/\psi D_s D_s}(q^2) f_{D_s}^2 f_{J/\psi} m^4_{D_s} m_{J/\psi}(p_\mu + p'_\mu)}{(m_c + m_s)^2 (p^2 - m_{D_s}^2)(p^2 - m_{J/\psi}^2)(p'^2 - m_{D_s}^2)  } + h. r. \,,\label{eq:fenomvnoff} \\
\Gamma^{phen (D_s)}_\mu(p,p') &=& \frac{g^{(D_s)}_{J/\psi D_s D_s}(q^2) f_{D_s}^2 f_{J/\psi} m^4_{D_s} m_{J/\psi}\left ( -\frac{q^2 - p^2 - p'^2}{m^2_{J/\psi}} p_\mu - 2 p'_\mu \right )}{(m_c + m_s)^2 (p^2 - m_{J/\psi}^2)(q^2 - m_{D_s}^2)(p'^2 - m_{D_s}^2)  } + h. r. \,,\label{eq:fenomdsoff}
\end{eqnarray}
where $h.r.$ stands for the contributions of higher resonances and continuum states. $g^{(M)}_{J/\psi D_s D_s}(q^2)$ is the form factor of the $J/\psi D_s D_s$ vertex with meson $M$ off-shell ($M=J/\psi, D_s$). 

\subsection{The OPE side}

The OPE side is obtained from Eqs.~(\ref{eq:pivoff}) and (\ref{eq:pidsoff}) writing the meson interpolating currents in terms of the quark fields. The currents used in this work are $j_\mu^{J/\psi} = \bar{c} \gamma_\mu c$, and $j_5^{D_s^-} = i \bar{c} \gamma_5 s$. 

\begin{figure}[ht]
  \includegraphics[width=450px]{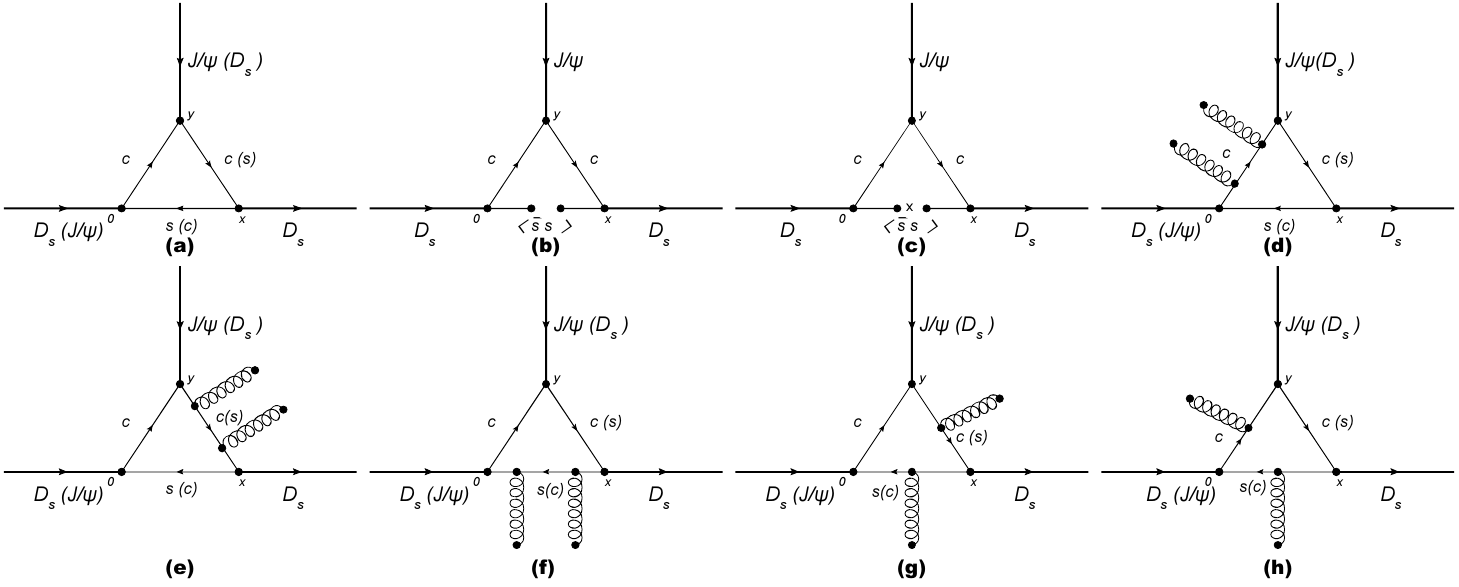}\\
	\includegraphics[width=450px]{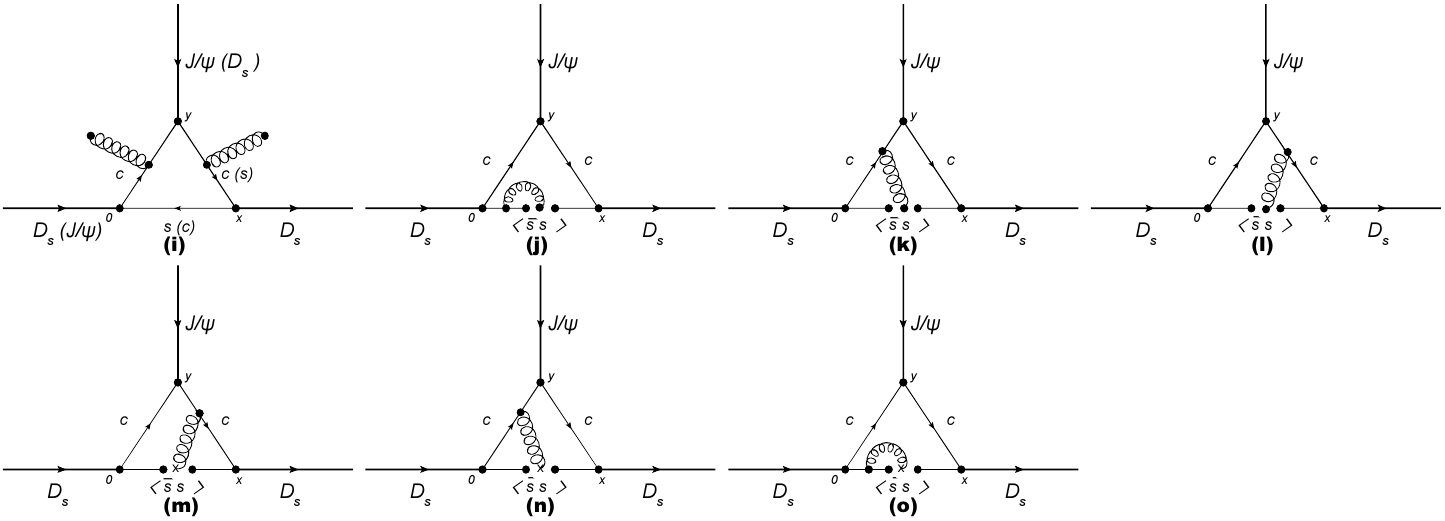}
  \caption{  \label{fig:diagrams}Contributing diagrams to the OPE side for $J/\psi \; (D_s)$ off-shell. }
\end{figure}

By construction, the OPE side is given by an expansion of the correlation function known as Wilson's Operator Product Expansion, which is dominated by a perturbative term ($\Gamma^{pert(M)}_\mu$) and followed by non-perturbative contributions ($\Gamma^{npert(M)}_\mu$):
\begin{align}
\Gamma^{OPE(M)}_\mu(p,p') = \Gamma^{pert(M)}_\mu + \Gamma^{npert(M)}_\mu \,.
\label{eq:piladodaqcdgeral}
\end{align}
When calculating form factors, the expansion usually exhibits a rapid convergence and then it can be truncated after a few terms. In this work, we consider non-perturbative contributions up to the mixed quark-gluon condensate, and we write: 
\begin{align}
\Gamma^{npert}_\mu  = \Gamma^{\langle\bar{q}q\rangle}_\mu + \Gamma^{m_q\langle\bar{q}q\rangle}_\mu + \Gamma^{\langle g^2 G^2\rangle}_\mu + \Gamma^{\langle \bar{q}g\sigma G q \rangle}_\mu + \Gamma^{m_q\langle \bar{q}g\sigma G q \rangle}_\mu\,.
\label{eq:nonpertcontrib}
\end{align}
All the relevant contributions are shown in Figure~\ref{fig:diagrams}. Note that only the $J/\psi$ off-shell case has contributions from all non-perturbative terms of Eq.~(\ref{eq:nonpertcontrib}). As a consequence of the double Borel transform used, all the non-perturbative terms, except the gluon condensates (Fig.~\ref{fig:diagrams}(d-i)), are suppressed in the $D_s$ off-shell case. This has been taken into account in Figure~\ref{fig:diagrams}, nevertheless, these contributions were omitted from the beginning.

Within the dispersion relations formalism, the perturbative term for a given meson $M$ off-shell (Fig.~\ref{fig:diagrams}(a)) can be written in the following form:
\begin{align}
\Gamma^{pert (M)}_\mu(p,p') = - \frac{1}{4\pi^2} \int^\infty_0 \int^\infty_0 \frac{\rho_\mu^{pert (M)}(s, u, t)}{(s-p^2)(u-p'^2)} ds du\,.
\label{eq:pertgeral}
\end{align}
where $\rho_\mu^{pert (M)}(s, u, t)$ is the spectral density of the perturbative term. After applying the Cutkosky's rules and using Lorentz symmetry, we have the following expression:
\begin{align}
\rho_\mu^{pert (M)}(s,u,t) = \frac{3}{2\sqrt{\lambda}}\left [ F_p^{(M)}(s,u,t)p_\mu +  F_{p'}^{(M)}(s,u,t)p'_\mu \right ] \,,
\label{dspec}
\end{align}
where $\lambda = (u + s - t)^2 - 4us$ and $F_p^{(M)}$ and $F_{p'}^{(M)}$ are invariant amplitudes. For the cases studied here, these invariant amplitudes can be written as:
\begin{eqnarray}
F_p^{(J/\psi)} &=& u(A - 1) - A(t-s) - (2A - 1)(m_s-m_c)^2\,,\\
F_{p'}^{(J/\psi)} &=& s(B - 1) + B(u - t) + (1 - 2B)(m_s-m_c)^2\,,\\
F_p^{(D_s)} &=& u(A + 1) + A(t-s) - (1 + 2A)(m_s-m_c)^2 \,,\\
F_{p'}^{(D_s)} &=& -s(B + 1) + B(u + t) - 2B(m_s-m_c)^2\,,
\end{eqnarray}
where 
\begin{align}
&A = \left [ \frac{\bar{k}_0}{\sqrt{s}} - \frac{p'_0 \overline{|\vec{k}|} \overline{\cos\theta}}{|\vec{p'}|\sqrt{s}} \right ]\,,
\;\;\;\;\;\;\;\;
B = \frac{\overline{|\vec{k}|} \overline{\cos\theta}}{|\vec{p'}|} \,,
\;\;\;\;\;\;\;\;
\overline{\cos\theta} = \frac{2p'_0\bar{k}_0 - u + (-1)^{(\epsilon)}(m_s^2 - m_c^2) }{2|\vec{p'}|\overline{|\vec{k}|}} \,,
\nonumber \\
&\overline{|\vec{k}|} = \sqrt{\bar{k}_0^2 - \epsilon m_s^2 - (1-\epsilon)m_c^2}\,,
\;\;\;\;\;\;\;\;
\bar{k}_0 = \frac{s + \epsilon(m_s^2-m_c^2)}{2\sqrt{s}}\,,
\nonumber \\
&p'_0 = \frac{s+u-t}{2\sqrt{s}}\,,\;\;\;\;\;\;\;\;|\vec{p'}| = \frac{\sqrt{\lambda}}{2\sqrt{s}}\,, \nonumber
\end{align}
and $\epsilon = 0(1)$ for $D_s(J/\psi)$ off-shell. The quantities $\bar{k}_0$, $\overline{|\vec{k}|}$, and 
$\overline{\cos\theta}$ are the centers of the $\delta$-functions appearing in the Cutkosky's rules and $s=p^2$, $u=p'^2$, and $t=q^2$ are the Mandelstam variables.

The first non-perturbative contribution to the correlation function comes from the strange quark condensate, diagram (b) in Fig.~\ref{fig:diagrams}, which is given by: 
\begin{align}
\Gamma^{\langle\bar{s}s\rangle (J/\psi)}_\mu = \frac{-m_c \langle\bar{s}s\rangle[p_\mu + p'_\mu ]}{(p^2 - m^2_c)(p'^2 - m_c^2)}\,.
\label{eq:contribcond}
\end{align}
Diagram (c) in Fig.~\ref{fig:diagrams} represents the strange mass-quark condensate, numerically less important, given by:
\begin{align}
\Gamma^{m_s\langle\bar{s}s\rangle (J/\psi)}_\mu = \frac{m_s\langle\bar{s}s\rangle \left ( 2m_c^2(p^2+{p'}^2)-p^2{p'}^2 - 3m_c^4 \right )[p_\mu + p'_\mu ]  }{2\left (p^2 - m_c^2 \right )^2 \left ({p'}^2 - m_c^2 \right )^2}\,.
\end{align}

Contributions coming from the charm-quark condensate are very small and can be safely ignored. 
Full expressions for the contributions coming from gluon condensates ($\langle g^2 G^2 \rangle$, Fig.~\ref{fig:diagrams}(d-i)) and mixed quark-gluon condensates ($\langle \bar{s}g\sigma \cdot G s \rangle$,  Fig.~\ref{fig:diagrams}(j-o)) for the case with $J/\psi$ off-shell can be found in Appendix A.

\subsection{The sum rule}

The sum rule is obtained after invoking the quark-hadron duality, which allows us to make the
connection between the phenomenological and OPE sides: 

\begin{equation}
{\cal BB}\left[\Gamma_\mu^{OPE(M)}\right](M,M')={\cal BB}\left[\Gamma_\mu^{phen(M)}\right](M,M')\,, \label{qhb}
\end{equation}
where a double Borel transform ($\cal BB$) was performed \cite{Khodjamirian:2002pka,Colangelo:2000dp}. The $\cal BB$ also implies in the variable transformations $P^2 = - p^2 \to M^2$ and $P'^2 = -p'^2 \to M'^2$, where $M$ and $M'$ are the Borel masses. 

Next, it is necessary to eliminate the $h.r.$ terms appearing in the phenomenological side in Eqs.~(\ref{eq:fenomvnoff}) and  (\ref{eq:fenomdsoff}). This is done by introducing the continuum cutoffs $s_0$ and $u_0$ in the integrations of the OPE side. These cutoffs satisfies the relations $m_i^2 < s_0 < {m'}_i^2$ and $m_o^2 < u_0 < {m'}_o^2$,  where  $m_i$ and $m_o$ are the masses of the incoming and outcoming mesons respectively, and $m'$ is the mass of the first excited state of these mesons. The quark-hadron duality lets us identify the integrals from $s_0$ and $u_0$ to infinity in Eq.~(\ref{eq:pertgeral}) with the $h.r.$ terms in the phenomenological side Eqs.~(\ref{eq:fenomvnoff})-(\ref{eq:fenomdsoff}), canceling them out from the sum rule.

After performing these two steps,  Eq.~(\ref{qhb}) allows us to obtain expressions for the form factors, in the case of the $p'_\mu$ structure they are:
\begin{eqnarray}
\label{eq:rsqcdvn}
g_{J/\psi D_s D_s}^{(J/\psi )}(Q^2) &=& \frac{-\frac{3}{8\pi^2} \int^{s_0}_{s_{inf}} \int^{u_0}_{u_{inf}} \frac{1}{\sqrt{\lambda}}F_{p'}^{(J/\psi)} e^{-\frac{s}{M^2}}e^{-\frac{u}{M'^2}} ds du + \mathcal{B}\mathcal{B}\left [ \Gamma^{npert}_{{p'}} \right]}{\frac{f_{D_s}^2 f_{J/\psi } m^4_{D_s} m_{J/\psi }}{(m_c + m_s)^2(Q^2+m_{J/\psi }^2)}e^{-m_{D_s}^2/M^2}e^{-m_{D_s}^2/M'^2}}\,,
\\
\label{eq:rsqcdds} 
g_{J/\psi  D_s D_s}^{(D_s)}(Q^2) &=& \frac{-\frac{3}{8\pi^2} \int^{s_0}_{s_{inf}} \int^{u_0}_{u_{inf}} \frac{1}{\sqrt{\lambda}}F_{p'}^{(D_s)} e^{-\frac{s}{M^2}}e^{-\frac{u}{M'^2}} ds du + \mathcal{B}\mathcal{B}\left [ \Gamma^{\langle g^2 G^2\rangle}_{{p'}}  \right]}{\frac{2 f_{D_s}^2 f_{J/\psi } m^4_{D_s} m_{J/\psi }}{(m_c + m_s)^2(Q^2+m_{D_s}^2)} e^{-m_{J/\psi }^2/M^2}e^{-m_{D_s}^2/M'^2}}\,,
\end{eqnarray}
where the first one is for meson $J/\psi$ off-shell and the second one for meson $D_s$ off-shell.

\section{Results and Discussion}

Eqs.~(\ref{eq:rsqcdvn}) and (\ref{eq:rsqcdds}) show the two different form factors for the vertex. In order to minimize the uncertainties when  extrapolating the QCDSR results, it is required that both form factors lead to the same coupling constant (at $Q^2= m^2_{J/\psi}$ and $Q^2= m^2_{D_s}$ respectively) \cite{Bracco:2001pqq}. 
A good sum rule is obtained if a good stability is achieved as a function of the Borel masses $M$ and $M'$. The set of Borel masses giving a good sum rule is called ``Borel window''. This window can be defined imposing that the ratio of the pole contribution to the total correlation function is at least $50 \%$ bigger than the continuum contribution and that the perturbative term contributes with more than $50\%$ of the total.

The Borel masses $M$ and $M'$ were chosen to be related through $M'^2 = \frac{m_o^2}{m_i^2}M^2$,
where $m_o$ and $m_i$ are the meson masses related to the $p'$ and $p$ moment respectivelly. We have calculated the average value of the form factor, not only for the complete Borel mass window, but also for each value of $Q^2$ used.  The standard deviation is then used to automatize the analysis of the stability of the form factor with respect  to the Borel masses and continuum threshold parameters, ensuring a good stability in the Borel window and in the whole $Q^2$ interval.

Besides the values of the Borel masses and the continuum  thresholds, we also need the values of meson  and quark masses, decay constants and condensates (all of them at the renormalization scale energy of about 1 GeV). The meson masses used are $m_{D_s} = 1.968$ GeV and $m_{J/\psi} = 3.097$ GeV \cite{Nakamura:2010zzi}. The values and errors of the other used quantities are given in Table \ref{tab:errors}.

The continuum threshold parameters $s_0$ and $u_0$ (Eqs.~(\ref{eq:rsqcdvn}) and (\ref{eq:rsqcdds})) are defined as $s_0 = (m_i + \Delta_i)^2$ and $u_0 = (m_o + \Delta_o)^2$, where $\Delta_i$ and $\Delta_o$ were obtained from imposing the condition of the  most stable sum rule.  In order to include the pole and excluding the $h.r.$ contributions in the correlation functions for mesons $D_s$ and $J/\psi$ (Eqs.~(\ref{eq:fenomvnoff}) and (\ref{eq:fenomdsoff})), the values for $ \Delta_{J/\psi}$ and $\Delta_{D_s}$ is near the gap between the mass pole and the first mass resonance \cite{Nakamura:2010zzi,Badalian:2011tb}. 

\begin{figure}[ht]
  \includegraphics[width=240px]{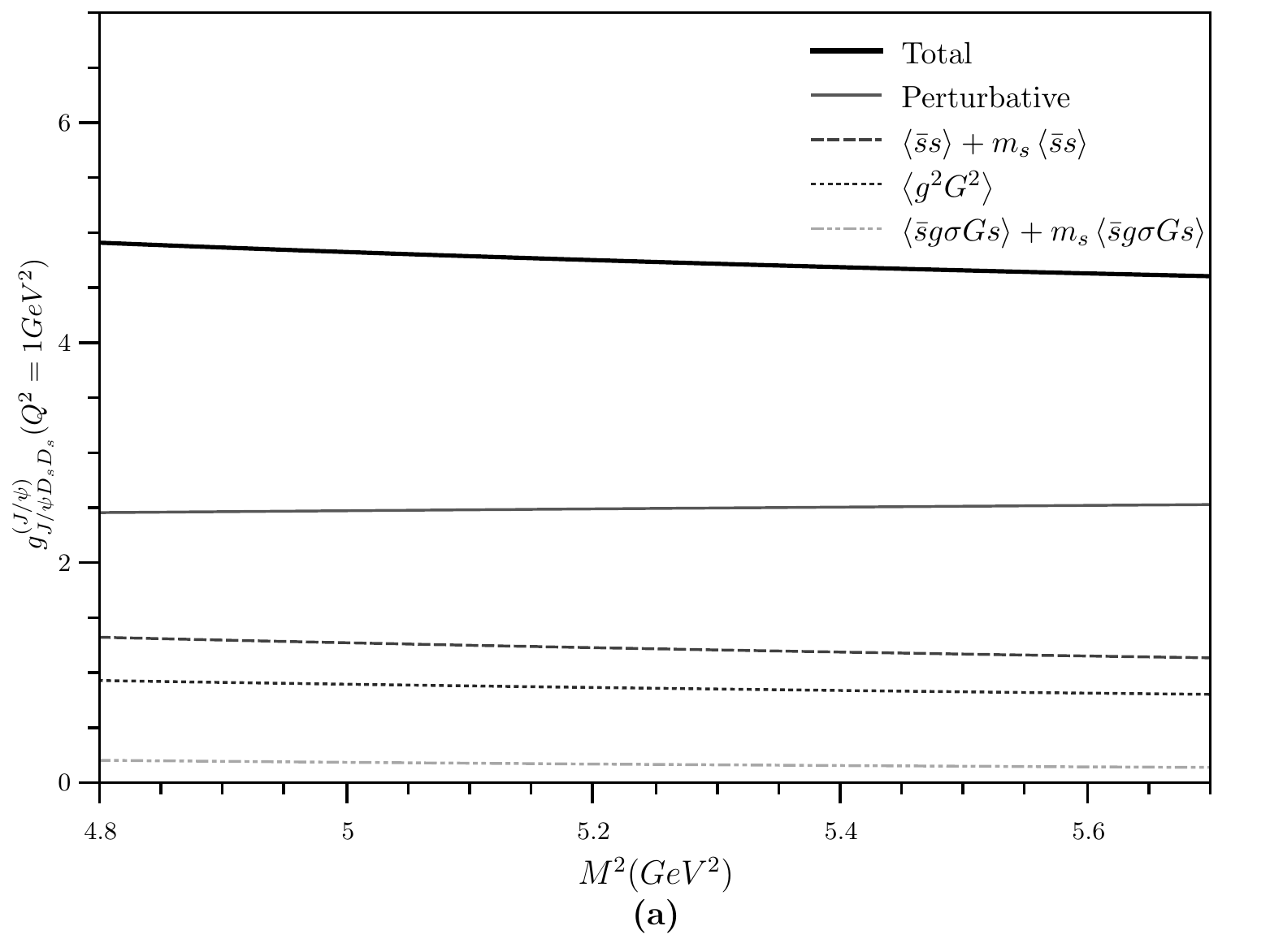}\includegraphics[width=240px]{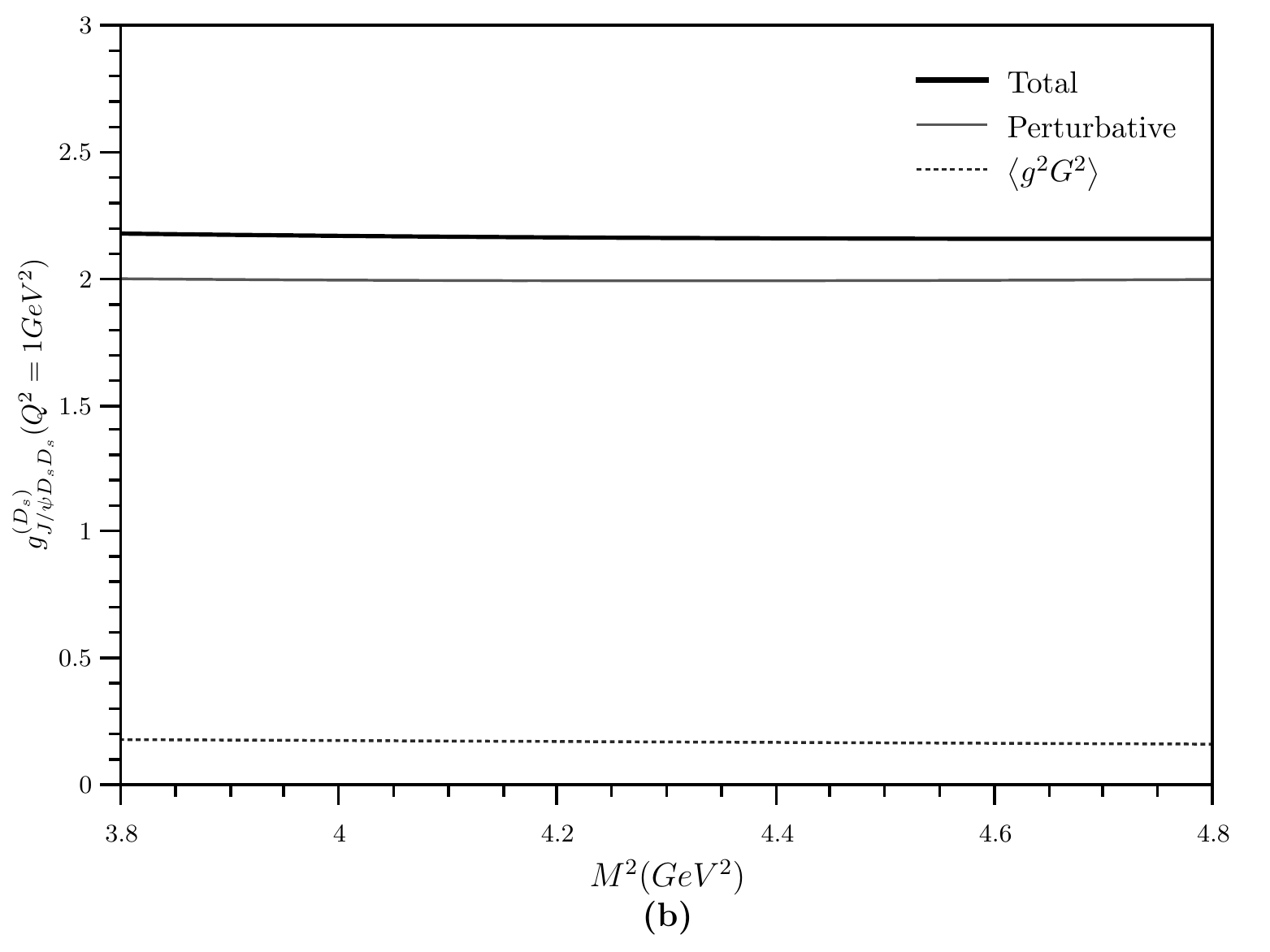}
  \caption{\label{fig:estabilidade}OPE contributions to the form factor for $J/\psi$ off-shell (panel \textbf{(a)}) and $D_s$ off-shell (panel \textbf{(b)}).}
\end{figure}

Our analysis shows that the best values are $\Delta_{D_s} = 0.6$ GeV and $\Delta_{J/\psi} = 0.5$ GeV, these figures lead us to a remarkably stable Borel window in both cases, as it can be seen in Fig.~\ref{fig:estabilidade}.
Also in Fig.~\ref{fig:estabilidade}, we can verify that the perturbative term is, in fact, the leading term in the OPE series, followed by the quark condensate (in the $J/\psi$ off-shell case) and the gluon condensates. The added contribution of $m_s \langle \bar{s}s \rangle$, $\langle \bar{s}g\sigma G s \rangle$ and $m_s \langle \bar{s}g\sigma G s \rangle$ terms is small, about $5\%$.

Regarding the choice of the Dirac structures used in the calculation, if the whole OPE series could be summed up, both structures, $p_\mu$ and $p'_\mu$ in Eq.~(\ref{dspec}), would lead to a valid sum rule. In a real calculation however, the OPE series has to be truncated at some order, and approximations are necessary to deal with the $h.r.$ terms in Eqs.~(\ref{eq:fenomvnoff}) and (\ref{eq:fenomdsoff}). Thus, it is not possible to use both structures in an equivalent way any more. For example, the structure $p_\mu$ for $D_s$ off-shell does not lead to a coupling constant compatible with the $J/\psi $ off-shell case, which is a necessary condition imposed upon the consistency of the calculation. 

For the $g_{J/\psi D_s D_s}^{(J/\psi)}(Q^2)$ form factor, the structure $p_\mu$ leads to results that are quite close to the ones obtained from using the $p'_\mu$ structure. The results presented in the following were calculated using the $p_\mu$ structure.

\begin{figure}[ht]
  \includegraphics[width=231px]{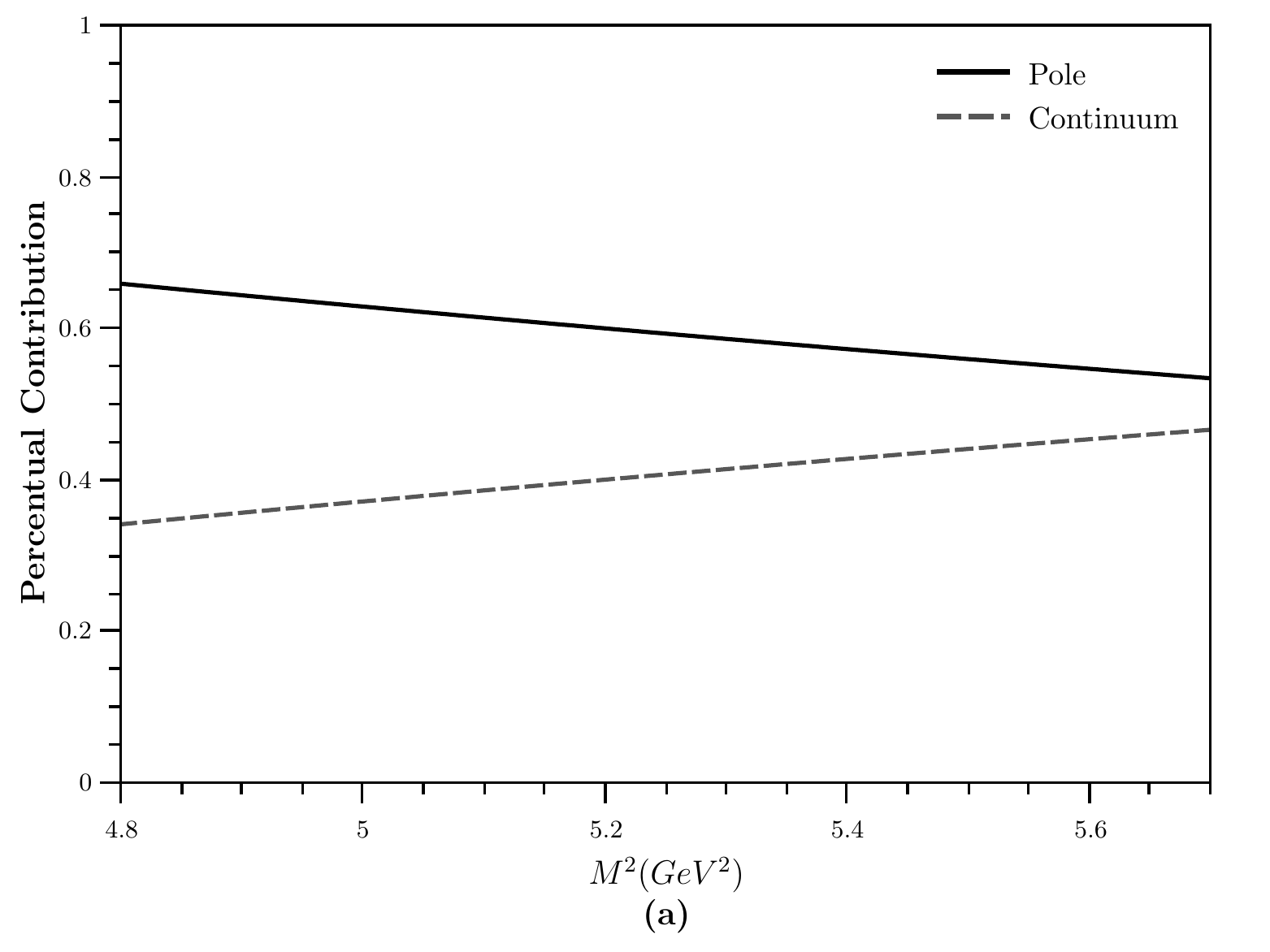}
  \includegraphics[width=231px]{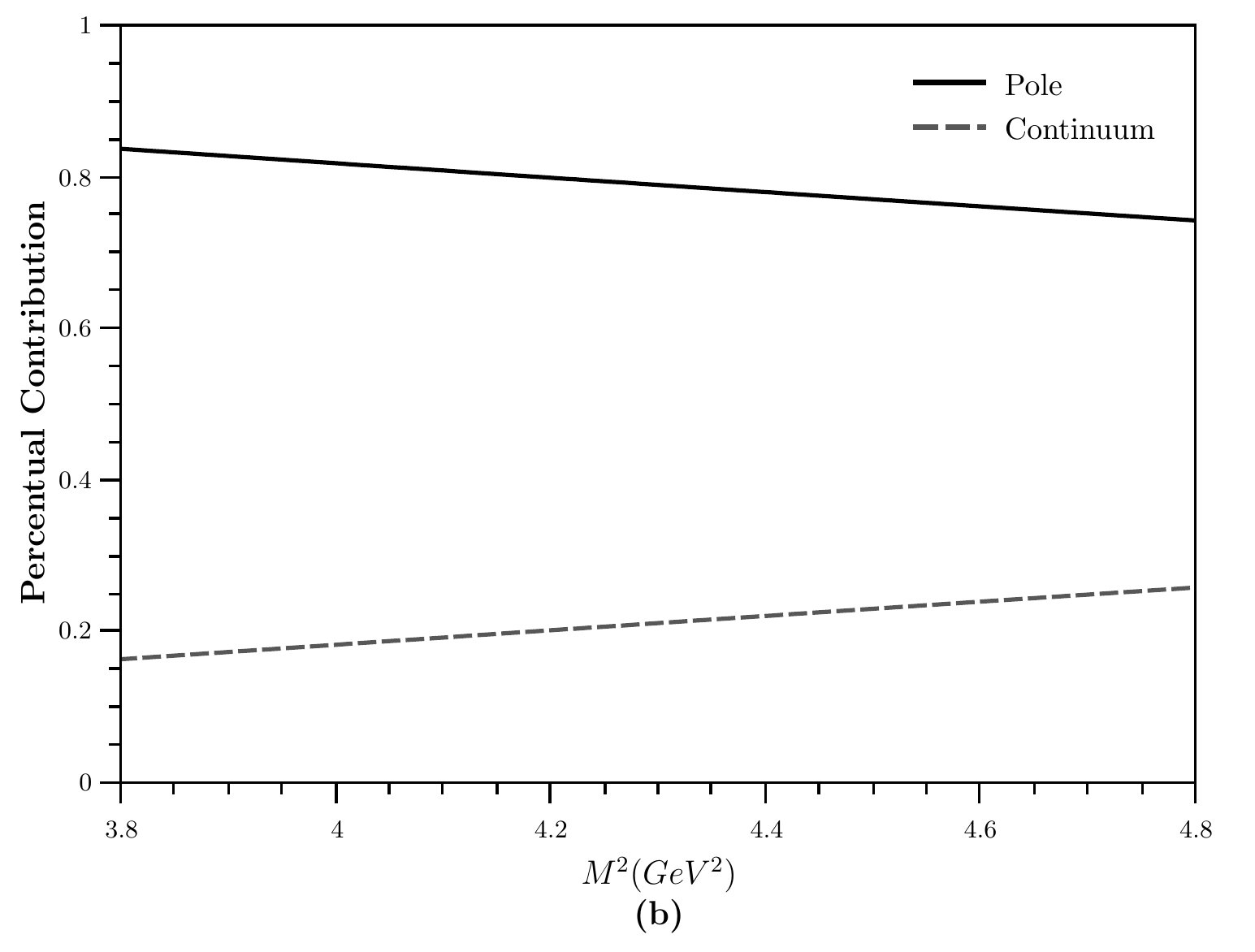}
  \caption{\label{fig:jpsidsdspolocont}Pole and continuum contributions for the $J/\psi$ off-shell (panel \textbf{(a)}) and for $D_s$ off-shell (panel \textbf{(b)}), both at $Q^2 = 1$ GeV$^2$.}
  \end{figure}

\begin{figure}[ht]
  \includegraphics[width=231px]{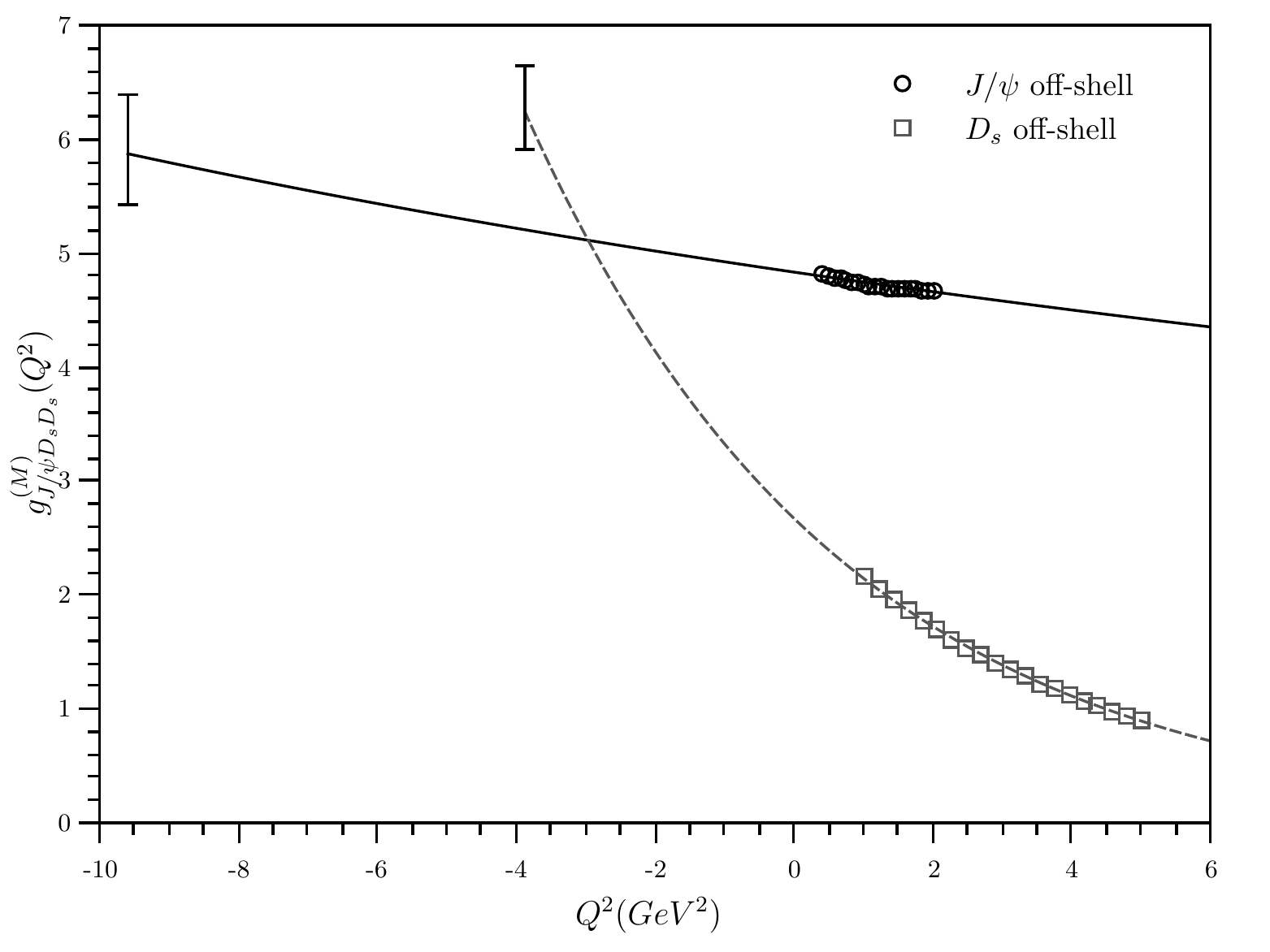}
  \caption{\label{fig:formfactors}Form factors of the $J/\psi D_s D_s$ vertex. Parametrizations are summarized in Table \ref{tab:results}. The coupling constants are represented by the points with the error bars.}
  \end{figure}

In Table~\ref{tab:results}, we present the $Q^2$ and $M^2$ Borel windows (first and second entries respectively) to obtain the form factors $g^{(M)}_{J/\psi D_sD_s}(Q^2)$ ($M=J/\psi,D_s$). Next entries are the parametrization function, the cut-off parameters $A$ and $B$ used in the parametrization and the coupling constant $g_{J/\psi D_sD_s}$ with its estimated error.  In Figure~\ref{fig:jpsidsdspolocont} it is shown how the Borel windows satisfy the dominance relations between the perturbative over the condensate contributions by at least 50\% of the total contribution. Regarding form factors, the function for the $J/\psi D_s D_s$ vertex with $J/\psi$ off-shell was well adjusted by a monopolar curve, while the $D_s$ off-shell case was well adjusted by an exponential curve (Fig.~\ref{fig:formfactors}).
We can also observe that the $B$ cut-off parameter is directly associated with the mass of the off-shell meson in the vertex. We say that a form factor is harder than another when its  curve as a function of $Q^2$ is flatter than the other. The hardest form factor has the largest  associated cutoff parameter. In our analysis we find that the form factor is harder if the off-shell meson is heavier.

\begin{table}[ht]
\caption{\label{tab:results}Parametrization of the form factors and numerical results for the coupling constant of this work. The calculation of $\sigma$ is explained in the text.}
\begin{ruledtabular}
\begin{tabular*}{\linewidth}{@{\extracolsep{\fill}} ccc}  
Quantity                      & $J/\psi$ off-shell & $D_s$ off-shell \\ 
\hline
$Q^2$ (GeV$^2$)            & [0.4, 2.0]    & [1.0, 5.0]   \\
$M^2$ (GeV$^2$)            & [4.8, 5.7]    & [3.8, 4.8]   \\
$g^{(M)}_{J/\psi D_sD_s}(Q^2)$ & $\frac{A}{B+Q^2}$ & $A e^{-Q^2/B}$\\
$A$                          & 262.2 GeV$^2$  & 2.674       \\
$B$ (GeV$^{2}$)             & 54.23       & 4.568        \\
${g^{(M)}_{J/\psi D_sD_s}}\pm\sigma$     & $5.87^{+0.52}_{-0.44}$          & $6.24^{+0.41}_{-0.36}$   \\
\end{tabular*}
\end{ruledtabular}
\end{table}

The coupling constant is defined as $\lim\limits_{Q^2 \to -m_M^2} g^{(M)}_{J/\psi D_s D_s}(Q^2)$ and should have a unique value, independently of what meson is off-shell. After extrapolating the form factors showed in Table~\ref{tab:results}, the resulting coupling constants agrees within the error uncertainties, as shown in Fig.~\ref{fig:formfactors}.

In Refs.~\cite{Rodrigues:2010ed} and \cite{CerqueiraJr2012130} we estimated the uncertainties of the coupling constants by studying their behavior when each parameter involved in the calculation was varied individually between their upper and lower limits . Experimental parameters have their own experimental errors (shown in Table~\ref{tab:errors}). Regarding the QCDSR parameters, we varied the thresholds $\Delta_s$ and $\Delta_u$ in $\pm 0.1$ GeV, the momentum $Q^2$ in $\pm 20\%$ and for the error, due to the uncertainty in the Borel mass $M^2$, we used the standard deviation of the average value of the form factor within the Borel window. Then, we calculated the mean value of all these contributions and their standard deviations. The same procedure was adopted in this work in order to estimate the error bars in the results. 

In Table~\ref{tab:errors}, we can see how the percentile variation of each parameter (first column) affects the final value of the coupling constants (second and third columns). The variation of most of the parameters (the decay constants, the Borel mass  and the strange quark masses, for example) has little impact on the value of the coupling constants. We have found,  however, that the error in the charm mass is the one with the biggest propagation over the value of the coupling constants. The sensitivity related to the Borel mass was already expected to be small, as we have obtained very stable Borel windows (Fig.~\ref{fig:estabilidade}). 

\begin{table}[ht]
\caption{\label{tab:errors}Percentage deviation of the coupling constants coming from the propagations of the error in each parameter.}
\begin{ruledtabular}
\begin{tabular*}{\linewidth}{@{\extracolsep{\fill}} ccc}
 Parameter                    & $\Delta g_{J/\psi D_s D_s}^{(J/\psi)}$(\%) & $\Delta g_{J/\psi D_s D_s}^{(D_s)}$(\%)\\ 
\hline
$f_{D_s} = 257.5 \pm 6.1$(MeV)\cite{Nakamura:2010zzi}     & 3.87  & 3.87   \\
$f_{J/\psi} = 416 \pm 6 $(MeV)\cite{Nakamura:2010zzi}     & 1.18  & 1.18 \\
$m_c = 1.27^{+0.07}_{-0.09}$ (GeV)\cite{Nakamura:2010zzi} & 14.8  & 14.6 \\
$m_s = 101^{+29}_{-21}$ (MeV)\cite{Nakamura:2010zzi}      & 3.33  & 3.66 \\ 
$M^2$ (GeV$^2$)\footnotemark[1]    & 2.81  & 5.05 \\
$\Delta_s \pm 0.1$(GeV),$\Delta_u \pm 0.1$(GeV)    & 10.7  & 2.55  \\
$Q^2 \pm 20\%$ (GeV$^2$)\footnotemark[1]           & 4.17  & 2.13  \\
$\langle \bar{s}s\rangle = -(290 \pm 15)^3$ (MeV$^3$)\cite{PhysRevD.87.034503} & 1.74 & - \\
$\langle g^2 G^2 \rangle = 0.88\pm 0.16$(GeV$^4$)\cite{Narison2012412} & 0.90 & 2.01\\
$\langle \bar{s}g\sigma \cdot G s \rangle = (0.8\pm 0.2)\langle \bar{s}s\rangle$(GeV$^5$)\cite{Ioffe2006232} & 1.41 & - \\
Fitting parameters  ($A$ and $B$)      & 13.5  & 0.47 \\
\end{tabular*}
\footnotetext[1]{The intervals for these quantities are those of Table~\ref{tab:results}.}
\end{ruledtabular}
\end{table}

Using the results presented in Table~\ref{tab:results} and averaging with the results coming from the $p'_\mu$ structure, we obtain the final result for the coupling constant:
\begin{align}
g_{J/\psi D_s D_s} = 5.98^{+0.67}_{-0.58}
\label{eq:ctejpsidsdsd}
\end{align}

\section{Conclusions}
In this work we have calculated the strong form factors and coupling constant of the $J/\psi D_s D_s$ vertex using the QCDSR technique. We have alternately considered mesons $J/\psi$ and $D_s$ off-shell together with non-perturbative contributions up to the mixed quark-gluon condensate. Our results come from a good QCDSR, which means not only a good pole/continuum dominance, but also a perturbative contribution that is greater than condensate contributions and a very stable Borel window for the whole $Q^2$ studied interval. The coupling constant with  $D_s$ off-shell was compatible with the coupling constant obtained using $J/\psi$ off-shell within the error bars. We have observed that the form factor is harder when the heavier meson is off-shell. This is compatible with previous results of our group, when other processes were calculated. The errors obtained are about $10\%$, which is lower than the expected $20\%$ appearing usually in other QCDSR calculations. 

The result of Eq.~(\ref{eq:ctejpsidsdsd}) can be useful to compare our previous QCDSR results, if we use the relations between coupling constant derived by $SU(3)$ broken symmetry. For example, invoking the SU(3) broken symmetry \cite{Liu:2006dq}, it is expected the relation $g_{J/\psi D_s D_s} = g_{J/\psi D D}$.  Using the value $g_{J/\psi D D} =  5.8 \pm 0.8$ from \cite{Matheus:2005yu}, we can see that this relation is maintained. Still within the SU(3) scheme, the relation $g_{J/\psi D_s D_s} = g_{J/\psi D^* D^*}$ should be valid \cite{Bracco:2011pg,Liu:2006dq}. Comparing our result for $g_{J/\psi D_s D_s}$ with $g_{J/\psi D^* D^*} = 6.2 \pm 0.9$ from \cite{Bracco:2004rx}, again, we obtain again compatible results. The $g_{J/\psi D_s D_s} = g_{J/\psi D D}$ SU(3) relation is violated by the order of $4\%$, almost the same value found between $g_{J\psi D^* D^*}$ and $g_{J/\psi D D}$ \cite{Bracco:2011pg}, which means that, in spite of the mass difference between the involved mesons, SU(3) is a reasonable symmetry for these vertices when using QCDSR.

\acknowledgments
M.E.B and B.O.R wants to thanks CNPq and CAPES respectively for the financial support.

\appendix
\label{appendix:remaningexpressions}
\section{}
Here we present full expressions for the contributions coming from the condensates $\langle g^2 G^2\rangle$ and $\langle \bar{q}g\sigma G q \rangle$ to the correlator of Eq.~(\ref{eq:nonpertcontrib}) in the $J/\psi$ off-shell case.

\begin{align}
\nonumber
 &\mathcal{B} \mathcal{B} \left[\Gamma^{\langle g^2G^2 \rangle}_{{p'}_\mu} \right ] =\\
 &\frac{\langle g^2G^2 \rangle}{96\pi^2} \int^\infty_{1/M^2} d\alpha e^{\frac{2 \alpha( m_c^2 - m_s^2 - \alpha m_c^2 M^2)}{\alpha M^2-1}+\frac{\alpha t M^2 - t}{2 M^2}}  \left (  F_{(d)} +  F_{(e)} +  F_{(f)} + F_{(g)} + F_{(h)} + F_{(i)} \right )
\end{align}

\begin{align}
\nonumber
&F_{(d)} = m_c(\alpha^3 m_c t M^4-4 \alpha^3 m_c^3 M^4+12 \alpha^2 m_c M^4-2 \alpha^2 m_c t M^2+12 \alpha m_s M^2+\alpha m_c t+4 \alpha m_c m_s^2\\
&-8 \alpha m_c^2 m_s+12 m_s+4 \alpha m_c^3-12 m_c)/16
\end{align}
\begin{align}
\nonumber
&F_{(e)} = m_c(\alpha^4 m_c t M^6-4 \alpha^4 m_c^3 M^6+2 \alpha^3 m_s t M^4-5 \alpha^3 m_c t M^4-8 \alpha^3 m_c^2 m_s M^4-4 \alpha^2 m_s M^4\\
\nonumber
&+12 \alpha^3 m_c^3 M^4+4 \alpha^2 m_c M^4-4 \alpha^2 m_s t M^2+7 \alpha^2 m_c t M^2+4 \alpha^2 m_c m_s^2 M^2+8 \alpha^2 m_c^2 m_s M^2+12 \alpha m_s\\ 
\nonumber
&M^2-12 \alpha^2 m_c^3 M^2-12 \alpha m_c M^2+2 \alpha m_s t-3 \alpha m_c t+8 \alpha m_s^3-12 \alpha m_c m_s^2-8 m_s+4 \alpha m_c^3+8 m_c)\\
&/(16(\alpha M^2-1))
\end{align}

\begin{align}
\nonumber
&F_{(f)} = m_s(\alpha^5 m_s t M^6-4 \alpha^5 m_c^2 m_s M^6-2 \alpha^4 m_s t M^4-12 \alpha^3 m_s M^4+6 \alpha^2 m_s M^4+24 \alpha^3 m_c M^4+6 m_s\\
\nonumber
&-12 \alpha^2 m_c M^4+\alpha^3 m_s t M^2+4 \alpha^3 m_s^3 M^2-8 \alpha^3 m_c m_s^2 M^2+4 \alpha^3 m_c^2 m_s M^2+18 \alpha^2 m_s M^2-12 \alpha m_s M^2\\
&-36 \alpha^2 m_c M^2+24 \alpha m_c M^2 -6 \alpha m_s+12 \alpha m_c-12 m_c)/(2\alpha^2M^2(\alpha M^2-1)^3)
\end{align}
\begin{align}
\nonumber
\nonumber
&F_{(g)} = (\alpha^2 t M^4-4 \alpha^2 m_c^2 M^4-2 \alpha t M^2+6 \alpha m_c m_s M^2+t+8 m_s^2-18 m_c m_s+4 m_c^2)\\
&/(4(1-\alpha M^2))
\end{align}
\begin{align}
\nonumber
\nonumber
&F_{(h)} = (\alpha^3 t M^6-4 \alpha^3 m_c^2 M^6+4 \alpha^2 M^6-2 \alpha^2 t M^4-6 \alpha^2 m_c m_s M^4-8 \alpha M^4+\alpha t M^2-12 \alpha m_c m_s M^2\\
&+12 \alpha m_c^2 M^2+4 M^2-4 m_s^2+18 m_c m_s-8 m_c^2)/(-4(1-\alpha M^2)^2)
\end{align}
\begin{align}
\nonumber
&F_{(i)} = (\alpha^4 t M^6-4 \alpha^4 m_c^2 M^6+4 \alpha^3 M^6-\alpha^3 t M^4+8 \alpha^3 m_c^2 M^4-4 \alpha^2 M^4-\alpha^2 t M^2+4 \alpha^2 m_s^2 M^2+\alpha t\\
&-24 \alpha^2 m_c m_s M^2-4 \alpha^2 m_c^2 M^2-8 \alpha M^2-12 \alpha m_s^2+24 \alpha m_c m_s+8)/(16\alpha (1-\alpha M^2))
\end{align}


\begin{align}
 \mathcal{B} \mathcal{B} \left[\Gamma^{\langle \bar{s}g\sigma G s \rangle}_{{p'}_\mu} \right ] = \frac{m_c\langle \bar{s}g\sigma G s \rangle }{4M^4}e^{-2\frac{m_c^2}{M^2}} \left (3M^2 - t + 4m_c^2 \right )
\end{align}

\begin{align}
 \mathcal{B} \mathcal{B} \left[\Gamma^{m_s\langle \bar{s}g\sigma G s \rangle}_{{p'}_\mu} \right ] = \frac{m_s\langle \bar{s}g\sigma G s \rangle}{24M^6}e^{-2\frac{m_c^2}{M^2}} \left (48M^4 + 2tM^2 - 9m_c^2M^2 + 16m_c^4 \right )
\end{align}



\end{document}